# Quantum sensing of acceleration and rotation by interfering magnetically-launched atoms

Clément Salducci,[1]* Yannick Bidel,[1]* Malo Cadoret,[1,2] Sarah Darmon,[1] Nassim Zahzam,[1] Alexis Bonnin,[1] Sylvain Schwartz,[1] Cédric Blanchard,[1] Alexandre Bresson[1]

1 DPHY, ONERA, Université Paris-Saclay, F-91123 Palaiseau, France
2 LCM-CNAM, 61 rue de Landy, 93210, La Plaine Saint Denis, France
*Corresponding authors: clement.salducci@onera.fr, yannick.bidel@onera.fr

September 11th, 2024

**Accurate measurement of inertial quantities is essential in geophysics, geodesy, fundamental physics and navigation. For instance, inertial navigation systems require stable inertial sensors to compute the position and attitude of the carrier. Here, we present an architecture for a compact cold-atom accelerometer-gyroscope based on a magnetically launched atom interferometer. Characterizing the launching technique, we demonstrate 700 ppm gyroscope scale factor stability over one day, while acceleration and rotation rate bias stabilities of $7 \times 10^{-7}$ m/s² and $4 \times 10^{-7}$ rad/s are reached after two days of integration of the cold-atom sensor. Hybridizing it with a classical accelerometer and gyroscope, we correct their drift and bias to achieve respective 100-fold and 3-fold increase on the stability of the hybridized sensor compared to the classical ones. Compared to state-of-the-art atomic gyroscope, the simplicity and scalability of our launching technique make this architecture easily extendable to a compact full six-axis inertial measurement unit, providing a pathway towards autonomous positioning and orientation using cold-atom sensors.**

## Introduction

Light-pulse atom interferometry is based on the coherent manipulation of matter waves where atoms inside a vacuum chamber serve as a perfect proof mass in free-fall, delivering absolute and stable inertial measurements. Over the past decades, these systems have demonstrated in laboratory environment exceptionally accurate and sensitive measurements of importance in fundamental physics, such as measurements of gravity (*1*, *2*), rotation rates (*3*), (*4*), fundamental constants (*5–7*), tests of general relativity (*8*) and search for new forces (*9*, *10*). On a separate front, strong efforts of ruggedization and miniaturization have been made to operate these systems outside of the laboratory (*11*), where rough environmental conditions have to be balanced despite the inherent reduced sampling rate and dynamic range of atomic inertial sensors compared to classical ones. Up to now, for field applications, only vertical atom accelerometers have been demonstrated, either in static conditions on ground (*12*, *13*), (*14*), (*15*) or in shipborne (*16*, *17*), (*18*) and airborne (*19–21*) environments using hybridization techniques (*22*) consisting in fusioning the cold atom accelerometer output with the one of a high bandwidth classical accelerometer, thus allowing to provide a continuous bias-free acceleration measurement.

Current classical inertial measurement units comprise three accelerometers and gyroscopes which deliver greatly resolved but biased measurements that accumulate over time, leading to uncertainties on the position and the attitude of the carrier that comes mostly from the gyroscopes limited performances (*23*). Atomic inertial sensors' inherent stability makes it a promising technology that could tackle these issues, benefiting to many GNSS-denied applications such as inertial navigation (*23*) and tunnel drilling (*24*), as well as satellite orientation for space gravity missions (*25*) or geophysics through vector gravimetry mapping (*26*). Its development is however still hindered by significant physics and engineering challenges as very few atom gyroscopes have been demonstrated so far and mostly in large meter-scale apparatus (*27, 28*), (*29*), (*30*), (*31*) which are not compatible with field applications. Building a cold atom gyroscope requires to open a physical area in the interferometer making its implementation in multi-axis atom interferometers more complicated (*32–34*), (*35*). So far, the only demonstrated atom interferometry setup with six-axis sensing consisted in a bulky laboratory-based experiment operating in static conditions, and utilizing two parabolically launched atom clouds and a complex combination of separate interferometry setups (*36*).

In this work, we report on a novel atom launching technique studied within the framework of a dual cold-atom accelerometer-gyroscope based on light-pulse atom interferometry. Harnessing the Stern-Gerlach effect, the cold-atom cloud is launched horizontally at a velocity of 8.2 cm/s with a magnetic field gradient pulse generated by two coils in anti-Helmholtz. The launch velocity is characterized precisely using Raman spectroscopy, achieving a stability of the atomic gyroscope's scale factor of 700 ppm. Additionally, we demonstrate the first quantum cold-atom accelerometer-gyroscope hybridized with both a classical accelerometer and a gyroscope. While until now hybridization between quantum and classical sensors addressed only accelerometers, even though in multi-axis configuration (*37*), we demonstrate the correction of both the drift and bias of a force-balanced accelerometer) and a Coriolis vibrating gyroscope at the same time, thus improving the long-term stability of both sensors. The hybrid sensor offers high-bandwidth measurements of acceleration and rotation rate with a short term sensitivity of $1.2 \times 10^{-6}$ m/s²/$\sqrt{\text{Hz}}$ and $1.8 \times 10^{-6}$ rad/s/$\sqrt{\text{Hz}}$ provided by the classical sensors and a stability over two days of $7 \times 10^{-7}$ m/s² and $4 \times 10^{-7}$ rad/s provided by the atom sensor which corresponds to an improvement of respectively 100-fold and 3-fold compared to the classical sensors alone. Angular velocity measurements up to 100 times Earth rotation are also reported. This atom launching technique is compatible with a compact design and could easily be scaled up to a six-axis sensor, paving the way towards the development of a compact fully hybridized cold-atom inertial measurement unit where ultimately the atom interferometers would correct the classical ones and deliver a continuous drift-free measurement of both accelerations and rotations.

## Results
**Experimental setup**
*Operating principle of the cold-atom accelerometer-gyroscope*

The cold-atom accelerometer-gyroscope is based on a Mach-Zehnder light-pulse atom interferometer (*38*). The core of the experimental setup has been described in detail in (*39*) and is illustrated in Figure 1.A. In short, a cold $^{87}$Rb atom cloud is formed with a standard MOT (Magneto Optical Trap) configuration consisting of three mutually orthogonal pairs of counterpropagating laser beams intersecting at the center of a quadrupole magnetic field created by a pair of vertically aligned anti-Helmholtz coils. Compared to (*39*), an additional pair of magnetic coils aligned horizontally has been implemented to the setup, allowing to shift the atom cloud position and to launch the atoms (see Material and Methods). At the end of the atoms loading in the MOT, the atom cloud position is shifted horizontally by 7 mm. Then, after switching-off the MOT magnetic fields, the atoms are further cooled down to ~ 2 µK using polarization gradient cooling. The cooling light is gradually turned-off and the atoms are released to fall freely under gravity. The atom cloud is then launched along the opposite direction compared to the MOT's position shift (see Fig.1.B). A horizontal magnetic-field gradient pulse of 20 ms is applied to horizontally launch the atoms in the state $|F = 2, m_F = 1\rangle$ at a velocity $v_l = 8.2$ cm.s$^{-1}$ to match the width of the Raman beam. It is followed by a state selection that allows to prepare the atoms in the interferometer's initial magnetic insensitive state $|F = 1, m_F = 0\rangle$. The state selection is delayed by 25 ms in order to prevent Eddy currents due to the magnetic gradient to disturb the atom interferometer. While the atoms are free-falling, a combination of three laser pulses separated by $T = 40$ ms is applied to perform atom interferometry in a Mach-Zehnder configuration ($\pi/2 - \pi - \pi/2$) as depicted in Figure 1.B. The beam splitter ($\pi/2$) and mirror ($\pi$) pulses of the interferometer use counter-propagating two-photon stimulated Raman transitions between the $|F = 1, m_F = 0\rangle$ and $|F = 2, m_F = 0\rangle$ clock states of the rubidium atom. Fluorescence detection is then used to compute the proportion of atoms in the $F = 2$ state at the output of the interferometer, $P_2 = P_m - (C/2)\cos(\Delta\Phi)$, where $P_m$ is the mean value, $C$ the contrast and $\Delta\Phi$ the interferometer phase difference accumulated between the two paths.

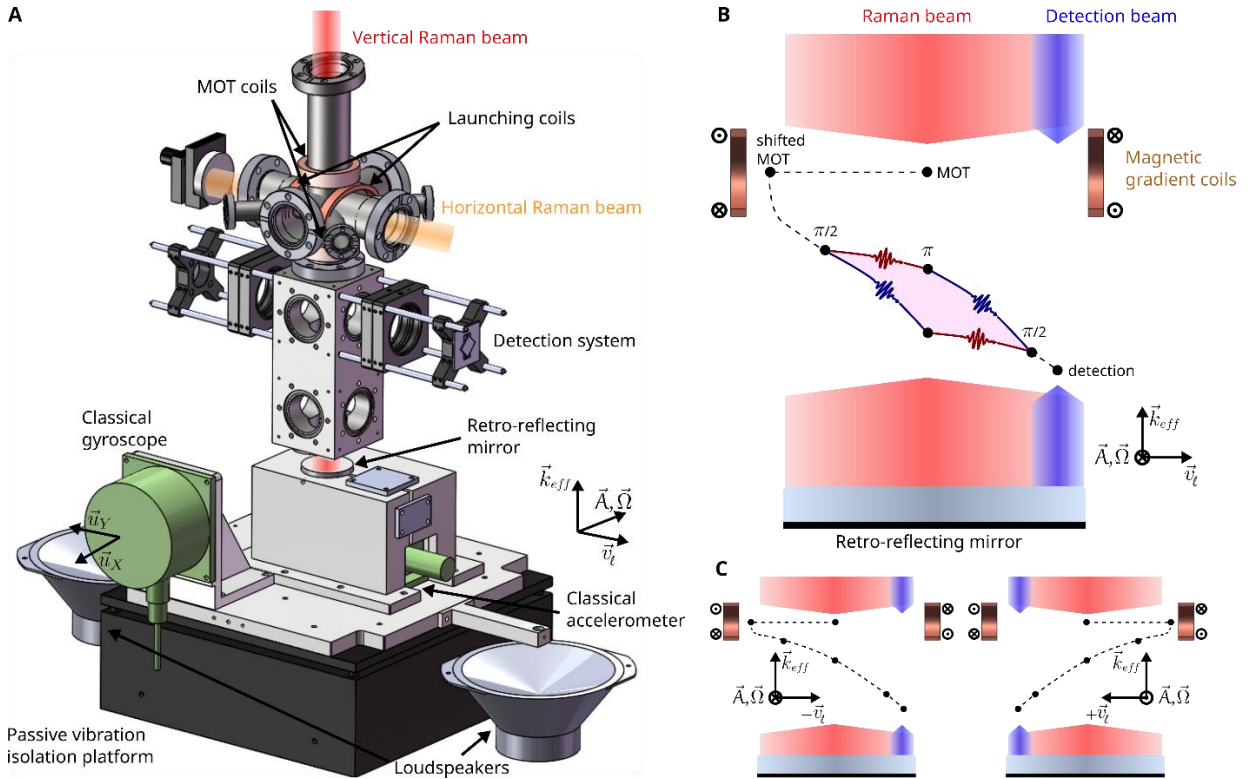

**Fig. 1. Principle of the experiment. (A)** Overview of the experimental apparatus where the atoms are laser cooled at the center of the 3D-cross and then accelerated thanks to the launching coils. The vertical Raman laser is used to perform the atom interferometer using two-photons counter-propagating transitions by being retro-reflected on a mirror that is set onto a passive vibration isolation platform. A classical accelerometer (Titan Nanometrics) attached to the mirror and a gyroscope (GI-CVG-U2200A Innalabs) are placed on the vibration isolation platform to record the accelerations and rotations experienced by the mirror which defines the inertial frame for the atoms. A pair of loudspeakers can be connected to the platform in order to perform dynamic rotation rate measurements along either $\vec{u}_X$ or $\vec{u}_Y$, the measurement axis of the classical gyroscope. Free-falling atoms that have performed the interferometry are detected 10 cm bellow the MOT region. **(B)** Mach-Zehnder interferometer diagram in the $(\vec{k}_{eff}, \vec{v}_l)$ plane (not to scale) exhibiting its two arms that enclose the physical area $\vec{A}$ (magenta), where the blue and red lines respectively account for the $F = 1$ and $F = 2$ states. The interferometer is sensitive to the vertical acceleration and to rotations orthogonal to the effective wave-vector of the Raman beams and the launching direction. **(C)** Atom interferometer's mid-point trajectory (not to scale) drawn for the negative

(left) and positive (right) launch velocities. To switch from one to the other, the sign of the current is switched in both coils and the detection beam (blue) is transversally shifted.

From this measurement the interferometer phase shift $\Delta\Phi$, and thereby the acceleration $\vec{a}$ and rotation rate $\vec{\Omega}$ of the atoms relative to the Raman retro-reflecting mirror (defined as the reference frame) can be determined. For atoms entering the interferometer with an initial velocity $\vec{v}_l$, the phase shift of the atom interferometer is (40):

$$\Delta\Phi = \left[\vec{k}_{eff}.\left(\vec{a} - 2\vec{\Omega} \times \vec{v}_l - \vec{\dot{\Omega}} \times \vec{r} - \vec{\Omega} \times \vec{\Omega} \times \vec{r}\right) - \alpha\right]T^2 \qquad (1)$$

where $\vec{k}_{eff}$ is the vertical effective wave vector of the two-photon counter-propagating Raman transition pointing orthogonally to the retro-reflecting mirror and $\alpha$ is the frequency chirp rate of the Raman lasers. This chirp rate adds a phase shift $\alpha T^2$ to the interferometer that, when properly tuned, exactly compensates for the phase shifts induced by the acceleration and rotation rate in average over the velocity distribution, enabling to measure the inertial quantities. The terms $2\vec{\Omega} \times \vec{v}_l$, $\vec{\dot{\Omega}} \times \vec{r}$ and $\vec{\Omega} \times \vec{\Omega} \times \vec{r}$ in equation (1) respectively represents the Coriolis, Euler and centrifugal accelerations experienced by the retro-reflecting mirror, where $\vec{r}$ is the position of the atoms with respect to the mirror and $\vec{\dot{\Omega}}$ is the angular acceleration.

Alternating the sign of the launch velocity $\pm\vec{v}_l$ allows to discriminate between the rotation and acceleration phase shifts (see eq. (1)). This is experimentally achieved by reversing the sign of the current in the pair of horizontal coils, which inverts both the direction of the MOT's displacement and the atom launch, while shifting horizontally the detection beam (see Fig. 1.C. and Material and Methods). The $+\vec{v}_l$ and $-\vec{v}_l$ atom interferometers are set to be symmetric from one to each other and to be centered on the Raman beam allowing to cancel most of the systematic effects. Additionally, the k-reversal technique is applied for each interferometer in order to remove other systematic effects independent of the direction of the effective wave-vector (41).

*Launch velocity analysis*
The sensitivity of the atom interferometer to rotation rates scales linearly with the velocity of the atoms at its input $\vec{v}_l$. A pair of quasi anti-Helmholtz magnetic coils are wrapped directly onto the vacuum chamber allowing to create a magnetic field gradient $\vec{\nabla}B$ pointing horizontally along the East-West axis. After polarization gradient cooling, the atoms are equally distributed among the five magnetic Zeeman sub-levels $m_F = \{-2, -1, 0, +1, +2\}$ of the $|F = 2\rangle$ hyperfine ground state. We harness the Stern-Gerlach effect to launch the atoms horizontally, resulting in a Zeeman state dependent magnetic force $\vec{F}_l$:

$$\vec{F}_l = \mu_B m_F g_F \vec{\nabla}B \qquad (2)$$

where $\mu_B$ is the Bohr magneton and $g_F$ is the Landé factor of the rubidium ground state. The atom cloud is therefore spatially divided in five parts that travel with different horizontal velocities. Following the launch, a state-selective microwave pulse allows to transfer the launched atoms from the internal state $|F = 2, m_F = +1\rangle$ to $|F = 1, m_F = 0\rangle$, while the atoms remaining in the $F = 2$ state are cleared away with a push beam such that they do not enter the atom interferometer. Additionally, because this microwave transition is degenerated with the transition $|F = 2, m_F = 0\rangle \rightarrow |F = 1, m_F = +1\rangle$, all the non-launched atoms are not cleared by the push beam. However, they fall far from the detection zone, making their impact on the phase shift measurement negligible (see Material and Methods).

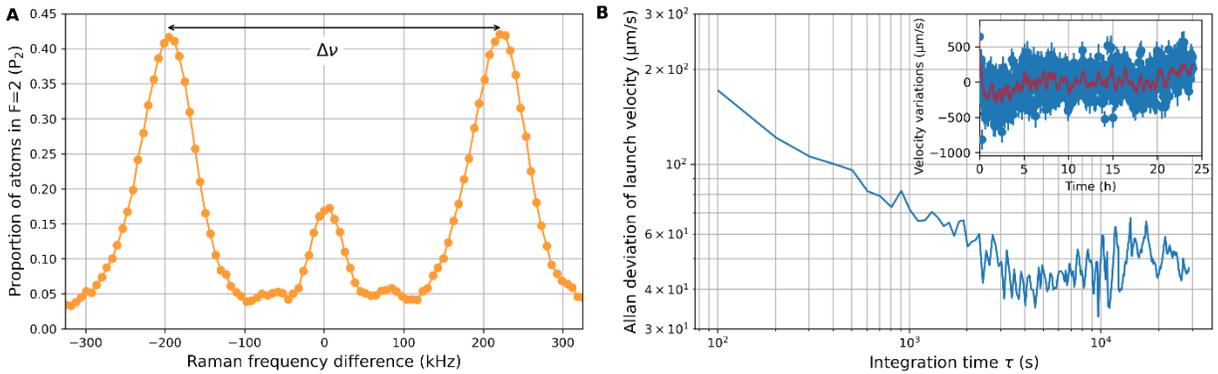

**Fig. 2. Measurement of the launch velocity. (A)** Typical launch velocity measurement spectrum where each point is obtained after one experimental cycle (see Materials and Methods) and plotted as a function of the Raman frequency difference subtracted from the hyperfine splitting of the $^{87}$Rb clock states. The launch velocity of the atoms is proportional to the frequency difference $\Delta\nu$ between the two counterpropagating transitions (side peaks) of effective wave-vector $\pm\vec{k}_{eff,h}$. Residual polarization defaults enable to see the two others degenerated copropagating transitions (middle peak). **(B)** A launch velocity measurement during one day is displayed in the inset, where each point (blue) is obtained from a Raman spectrum recorded in 100 s. The error bars on the individual measurements correspond to the statistical error on the frequency difference measurement. A moving average (red) is calculated for 1000 s segments. The Allan standard deviation of this time series measurement is plotted, exhibiting a velocity stability at a level of 60 µm/s rms over 24 hours.

The launch velocity is measured by Raman spectroscopy using the horizontal Raman laser beams (see Fig 1.A.). The resonance condition of the two-photon Raman transition depends on the velocity of the atoms and its alignment with the horizontal laser's effective wave-vector via the Doppler frequency term $\omega_D = \vec{k}_{eff,h} \cdot \vec{v}_l$ (*38*). In a retro-reflected configuration, two co-propagating and two counter-propagating transitions distinguished by the norm and the sign of their effective wave-vector are possible, resulting in Doppler frequency terms $\omega_D$ of different signs and amplitudes. Accurate launch velocity determination is made by measuring the frequency difference $\Delta\nu$ between the two counterpropagating transitions $\pm\vec{k}_{eff,h}$, $v_l = \pi\Delta\nu/k_{eff,h}$. Figure 2.A. shows a typical velocity measurement spectrum as a function of the Raman frequency difference. The resonance of each atomic transition might be shifted due to the coupling with other non-resonant atomic states (*41*) inducing systematic effects that can lead to significant errors on the velocity measurement. The correction from these systematic effects is detailed in the Materials and Methods.

Figure 2.B. illustrates a measurement of the launch velocity stability during one day corrected from the light shift effects. The Allan deviation of the launch velocity time series exhibits a velocity stability at a level of 60 µm/s (700 ppm) rms over 24 hours. The launch velocity stability limitation may arise both from an instability of the magnetic field gradient due to current fluctuations in the coils and a variation of the atom's mean velocity at the end of the polarization gradient cooling. Previous work on atomic gyroscopes where the atoms are launched with moving molasses (*42*) have reported a velocity stability of 30 µm/s over 1.5 hours of integration due to polarization instabilities. On the other hand, the coils are controlled in voltage, which can lead to electric current variations of the same order if their resistance

varies with the room temperature for instance. Both limits can be overcome by stabilizing the polarization of the cooling beams and the electric current flowing inside the launching coils.

## Stabilization of the classical sensors in static conditions
### *Data acquisition*
We performed a continuous acquisition of the atom interferometer over 44 hours, at a cycle rate of 2 Hz. The mid-fringe algorithm (*43*) is used to compute the value of $\alpha$ that cancels the phase shift $\Delta\Phi$ (see Equation (1)) every 2 shots by measuring at each side of a fringe (see Fig. 3). The signs of both the effective wave-vector $\pm\vec{k}_{eff}$ and the launch velocity $\pm\vec{v}_l$ are alternated to reject non-inertial systematic effects. Thus, the acceleration $a$ and rotation rate $\Omega$ measured by the atomic dual-axis sensor can both be computed every 4 seconds (8 shots) by combining the different values of $\alpha_{\pm k,\pm v}$ for the two signs of the effective Raman wave-vector $\pm\vec{k}_{eff}$ and the launch velocity $\pm\vec{v}_l$:

$$a = \frac{[(\alpha_{+k,-v} - \alpha_{-k,-v}) + (\alpha_{+k,+v} - \alpha_{-k,+v})]}{4k_{eff}}$$
$$\Omega = \frac{[(\alpha_{+k,-v} - \alpha_{-k,-v}) - (\alpha_{+k,+v} - \alpha_{-k,+v})]}{8v_l k_{eff}} - \frac{\Delta a_{vib}}{8v_l} \tag{3}$$

where $\Delta a_{vib}$ corresponds to the variation of acceleration between the two $-\vec{v}_l$ cycles and the two $+\vec{v}_l$ cycles (see Material and Methods). Compared to equation (1), the rotation rate is time-invariant ($\dot{\Omega} = 0$) in static conditions and the centrifugal term ($\Omega^2$) is rejected by the $\pm\vec{v}_l$ alternation as it does not depend on the sign of launch velocity.

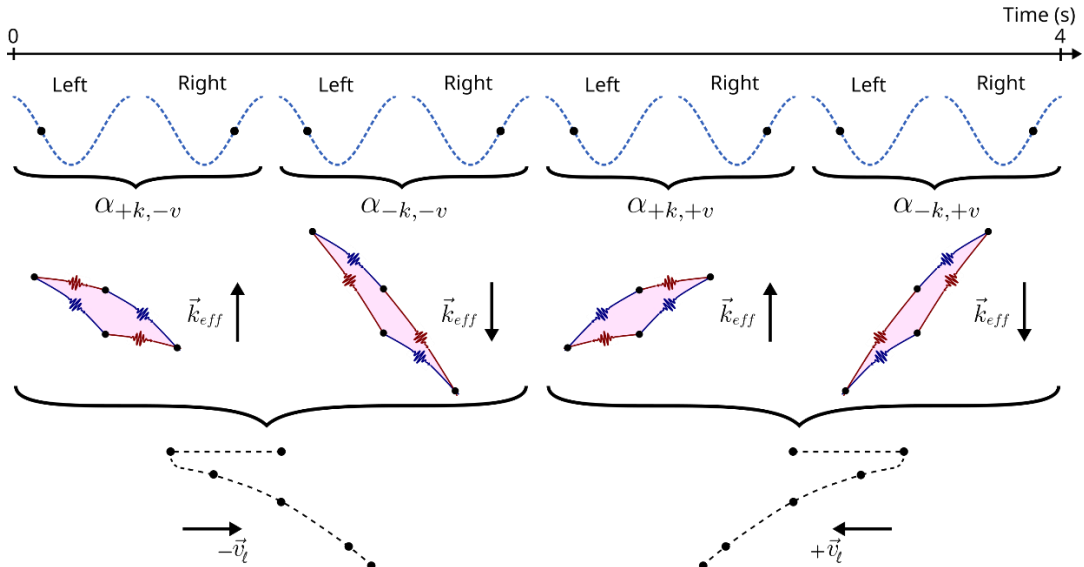

**Fig 3. Schematic timeline of one acceleration and rotation rate measurement.** The mid-fringe algorithm calculates a value of the atom interferometer phase shift every second (2 shots), and the sign of the effective Raman wave-vector $\pm\vec{k}_{eff}$ and the launch velocity $\pm\vec{v}_l$ are respectively alternated every 1 and 2 seconds (2 and 4 shots) to compute the four different phase shifts $\alpha_{\pm k,\pm v}$.

Atoms are launched approximately along the East-West axis such that the dual accelerometer-gyroscope measures the maximum of the Earth's rotation projected onto the latitude of Palaiseau (48.7° N). Both classical sensor outputs are also continuously acquired but only the average value of their output signal over one experimental cycle is recorded to avoid an unnecessary storage of data. All these raw data are post-processed to achieve a hybridized dual-sensor where the bias of both classical sensors is periodically corrected with the quantum dual accelerometer-gyroscope.

*Cold-atom accelerometer-gyroscope sensitivity*
Low sampling rate and non-continuous measurements make an atomic inertial sensor noise usually limited by the vibrations due to aliasing effects (*1*). In our dual-axis quantum sensor, the launch velocity direction is alternated in order to isolate the rotation from the acceleration measurement through a differential measurement. However, as the atom interferometers are not performed simultaneously but sequentially, the acceleration phase shift may vary between subsequent measurements and thus not being canceled. Consequently, uncompensated acceleration due to ground vibrations must be accounted for, adding the term $\Delta a_{vib}/2v_l$ to the rotation rate measurement. The force-balanced accelerometer records the vibrations during the atom interferometer, enabling to partially compensate for the extra-acceleration term in equation (3) and therefore to reduce the quantum sensor's noise, by a factor 5 for the data shown in this work (see Material and Methods).

Figure 4.A shows the Allan standard deviation of the acceleration measurement as defined in equation (3). For short times, it also integrates as $1/\sqrt{\tau}$ indicating white noise limited sensitivity of $3 \times 10^{-6}$ m/s²/$\sqrt{\text{Hz}}$ and drifts after 1000 s of integration to reach a plateau of $7 \times 10^{-7}$ m/s² after two days of integration. The Allan standard deviation of the rotation rate measurement corrected from vibrations is presented in Figure 4.B. It scales as $1/\sqrt{\tau}$ indicating white noise limited sensitivity of $1.1 \times 10^{-5}$ rad/s/$\sqrt{\text{Hz}}$ for short times. This value is a factor 3 above the detection noise limit and a factor 30 above the quantum projection noise limit of $3.5 \times 10^{-7}$ rad/s/$\sqrt{\text{Hz}}$. We estimate the short-term sensitivity of the rotation measurement to be limited by the residual ground vibration noise that has not been eliminated in the vibration removal process (see Material and Methods). After two days of integration, it reaches a plateau of $4 \times 10^{-7}$ rad/s.

*Bias correction of the classical sensors*
Classical (ie: non-quantum) inertial sensors are usually gifted by a lot of qualities that suit to field applications: low volume, continuous measurement and high dynamic range. However, they suffer from a lack of stability with time and the need to be calibrated. The Allan deviations of both classical inertial sensors displayed in Figure 4 exhibit major drifts of the bias of both sensors respectively after 50 s for the force-balanced accelerometer and 1000 s for the vibrating gyroscope.

The bias of both classical sensors is periodically corrected by the atomic dual accelerometer-gyroscope. This is done by implementing a feedback loop on the classical sensors' outputs:

$$\mu_n^{hyb} = \mu_n^{class} + b_n^{\mu}$$
$$b_n^{\mu} = b_{n-1}^{\mu} + G_{\mu}\{(\mu_n^{at} - \mu_n^{class}) - b_{n-1}^{\mu}\} \quad (4)$$

where $\mu_n$ can either be the acceleration $a$ or the rotation rate $\Omega$, *class*, *at* and *hyb* respectively indicate for the classical, atomic and hybridized outputs and $b_n^{\mu}$ are the classical sensor's biases estimated by the nth atomic measurement. The bias $b_n^{\mu}$ is computed using the last estimation of the bias $b_{n-1}^{\mu}$ and the term $(\mu_n^{at} - \mu_n^{class}) - b_{n-1}^{\mu}$ that can be identified as the error signal of the feedback loop with a gain $G_{\mu}$. Setting these gains to match with the crossing point of the atomic and classical sensors Allan deviations for each inertial quantity, the hybrid sensor embraces the advantages of both technologies, combining the highest sensitivities of the classical sensors and the stability of the quantum one. The Allan deviations of both hybridized signals are plotted in Figure 4, demonstrating a respective 100-fold and 3-fold improvements on the acceleration and rotation rates stabilities compared to both classical sensors operated alone.

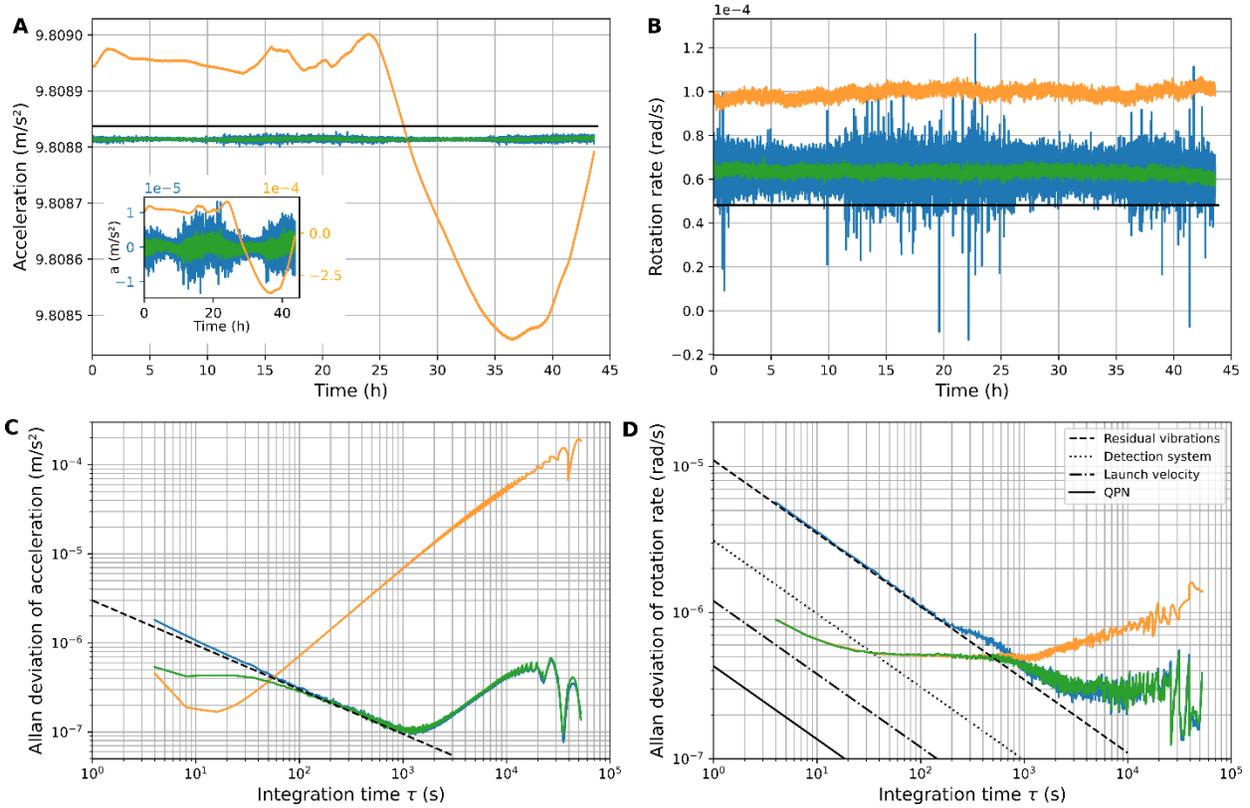

**Fig. 4. Correction of both classical sensors' bias.** Both acceleration **(A)** and rotation rate **(B)** temporal tracks are displayed for the classical (orange), quantum (blue) and hybridized (green) sensors over the 44 hours of measurement. Black solid lines represent the respective value of the gravity and Earth's rotation that should be measured by the atom interferometer. A dual y-axis plot is shown in the inset of **(A)** to resolve the noise of the signals. Respective Allan deviations are plotted in **(C)** and **(D)** with the $\tau^{-1/2}$ scaling (black-dashed) corresponding to the acceleration and rotation rate short term sensitivities of the quantum sensor. For the rotation rate measurement **(D)** we indicate the limits related to the detection noise (dotted) of $3.1 \times 10^{-6}$ rad/s/$\sqrt{\text{Hz}}$, the launch velocity noise (dot-dashed) of $1.2 \times 10^{-6}$ rad/s/$\sqrt{\text{Hz}}$ and the quantum projection noise (QPN) (solid) of $3.5 \times 10^{-7}$ rad/s/$\sqrt{\text{Hz}}$.

*Accuracy of the cold-atom accelerometer-gyroscope*
Although the hybridization algorithm allows to project the output value of the classical sensors onto the atomic ones, there remains a shift between both measured and expected values. The local gravity in the lab was measured using another atomic gravimeter to be 980 883.743(9) mGal (1 mGal = $10^{-5}$ m/s²) at the level of the mid-point of our atom interferometer, while we report here a value of 980 881.397(60) mGal. This difference of about 2.3 mGal is most likely due to a misalignment with the local verticality by a few mrad.

The projection of the Earth's rotation rate onto the measurement axis of our sensor is estimated to be $4.82(1) \times 10^{-5}$ rad/s. With our cold-atom accelerometer-gyroscope, we measure a value of $6.28(4) \times 10^{-5}$ rad/s, exhibiting a significant error of about 25% ($1.46 \times 10^{-5}$ rad/s) compared to the expected value. In our setup, we estimate that the main systematic effect comes from wave front distortions of the Raman beams that are not perfectly canceled due to the slight dissymmetry between opposite launch velocity configurations. We estimate that a dissymmetry of 1.2 mm between the two launching configurations, combined with a default of 1.9 rad peak to valley on the optical wave front, compatible with the quality of optics of $\lambda/6$ installed on our apparatus, could be responsible of the observed error (see Material and Methods). This error could therefore be mitigated by optimizing the symmetry and use optics of better qualities. Nevertheless, such a dissymmetry of 1.2 mm is quite large. It will be investigated in the future.

**Dynamic rotation rate measurements**

Our quantum accelerometer-gyroscope aims at addressing on-board applications where the environment is rougher than in the laboratory. In the following section, we present dynamic rotation rate measurements along the sensitive axis of our atomic gyroscope, where the retro-reflecting mirror is dynamically rotated with angular velocities up to a hundred times the Earth's rotation rate (ie: 4 mrad/s). This represents a first step towards operating the quantum sensor to rotation rates compatible with real environment.

*Data acquisition*

The simulation of a dynamic environment is made using a pair of loudspeakers fixed onto the vibration isolation table (see Fig. 1.A.) and that are operated in phase opposition. It produces two forces of opposite directions on each side of the table, making the upward board to oscillate periodically around its center of mass, which also corresponds to the center of rotation of the retro-reflecting mirror. The resulting rotation rate vector is of the form $\vec{\Omega}(t) = \Omega_d \cos(2\pi(t-T)/T_c + \varphi_0)\vec{u}_X$, where $T_c$ is the duration of one experimental cycle, $\vec{u}_X$ is the rotation rate vector's direction that is aligned with the classical gyroscope's measurement axis (see Fig. 1.A.) and $\varphi_0 = 0$ is set such that the atom interferometer is performed during a maximum portion of the sinusoidal function.

Rather than alternating the sign of the launch velocity every shot, here we have recorded the data by scanning the atomic fringes for both signs of the launch velocity $\pm \vec{v}_l$ during two separate days. The reason comes from the level of ground vibrations that was a lot higher than for the static study, due to the physical link of the mirror with the ground through the loudspeakers, which prevented from using the mid-fringe algorithm. For each launching direction $\pm \vec{v}_l$, the fringes are scanned by changing the chirp rate $\alpha$ from shot to shot (see Fig 5.A. and eq. (1)), and the amplitude of the rotation rate $\Omega_d$ is retrieved by comparing the shift of the fringe pattern to static conditions:

$$\Omega = \frac{\left(\alpha_{+k}^{\Omega_d} - \alpha_{-k}^{\Omega_d}\right) - \left(\alpha_{+k}^{\Omega_d=0} - \alpha_{-k}^{\Omega_d=0}\right)}{4 v_l k_{eff}} \quad (5)$$

where $\alpha_{\pm k}^{\Omega_d}$ is the Raman frequency chirp that cancels the total phase shift and are shown as the black star markers in Fig 5.A. Two fringes patterns are plotted for each sign of $\pm \vec{k}_{eff}$ enabling to remove systematic effects independent of the direction of the effective wave-vector. The vibration noise is subtracted in post-processing (see Material and Methods) for each shot of atomic fringes.

This configuration allows to study the quantum dual accelerometer-gyroscope's response in a dynamically simulated environment without using a dedicated rotation table. However, only the retro-reflecting mirror is rotated here which is not exactly equivalent to rotating the whole sensor (*44*) due to the variation of the norm of $\vec{k}_{eff}$. This adds a centrifugal-like term scaling in $\Omega^2$ which we have neglected here.

*Response to dynamic rotation rates*
The atom cloud's finite temperature of 2 µK makes the contrast of the fringes exponentially decrease with the rotation rate's amplitude because all the atoms do not experience the same Coriolis phase shift (see eq. (1)). This contrast decay is proportional to $e^{-2k_{eff}^2 \sigma_v^2 T^4 \Omega_d^2}$ (*45*), where $\sigma_v$ is the velocity dispersion associated to the temperature of the atom cloud and $T = 40$ ms, which limits the dynamic rotation range of the quantum gyroscope to 4 mrad/s. The contrast decreases as a function of the rotation amplitude $\Omega_d$ is plotted in Figure 5.A. It is fitted with a decaying exponential function giving a velocity dispersion equivalent to 1 µK in agreement with the estimated cloud temperature.

The response of the quantum sensor to rotation rate $\Omega$ is computed using equation (5) and shown in Fig. 5.B. for the two launching directions $\pm \vec{v}_l$. It exhibits a linear scaling of the atom interferometer's phase shift with the amplitude of rotation $\Omega_d$. The uncertainty bars are calculated as the quadratic mean of the statistical uncertainties $\sigma_{stat}$ and the stabilities of the bias $\sigma_{bias}$ and the scale factor $\sigma_{SF}$ for both gyroscopes, performing horizontal Raman spectroscopy at random times to estimate the scale factor stability of the atomic gyroscope. For dynamic rotation rate measurements, the scale factor instability becomes the largest source of error as it scales proportionally to the amplitude of the rotation rate. In particular, the velocity measurements associated to the $+\vec{v}_l$ plot have shown important velocity variations of 1.3 mm/s peak-to-peak, corresponding to a 2% variation of the scale factor. The rotation rate differences between the atomic and classical sensors plotted in the bottom of Fig. 5.B. indicate an agreement varying between <1% for $-\vec{v}_l$ and 5% for $+\vec{v}_l$. In equation (5), we have supposed the Euler and centrifugal accelerations to be negligible compared to the Coriolis one. This assessment might not be true for the Euler acceleration if the position of the atom cloud at the $\pi$-pulse is horizontally shifted by $d_0$ from the axis of rotation of the mirror and, in the meantime, $\varphi_0 \neq 0$. Estimating $d_0 \approx 1$ cm, the contribution of the Euler phase is of the order of 5% if $\varphi_0 \approx 0.02$ rad, which is a reasonable error given the experimental conditions. This result underlines the contribution of the Euler acceleration in the gyroscope's scale factor, that could possibly be significantly larger in on-board condition where $\varphi_0$ is not an adjustable parameter. This term could however be rejected by implementing the counter-propagating launching operation, that cancels the Euler acceleration if the two $\pm \vec{v}_l$ AI are perfectly symmetric and their symmetry axis is mixed with the vertical Raman beam axis.

The pair of loudspeakers can also be placed along the $\vec{u}_X$ axis to make the mirror rotate along the $\vec{u}_Y$ axis. If the $\pm \vec{v}_l$ atomic trajectories are tilted by an angle $\beta_\pm$ from $\vec{e}_y$, we should observe a shift of the atomic fringes by $\beta_\pm \Omega_d$ that is proportional to the rotation rate amplitude $\Omega_d$. We thereby estimate the mutual alignment of the opposing magnetically-launched trajectories to be $\beta_+ - \beta_- = (5 \pm 2)°$. This misalignment is not the cause of the 25% accuracy error in static conditions but shall be investigated in a further study. For each $\pm \vec{v}_l$ dataset of Fig. 5.B., the classical rotation rate is computed as $\Omega_{class,X} \cos(\beta_\pm) + \Omega_{class,Y} \sin(\beta_\pm)$ to take the misalignments between the two gyroscopes into account.

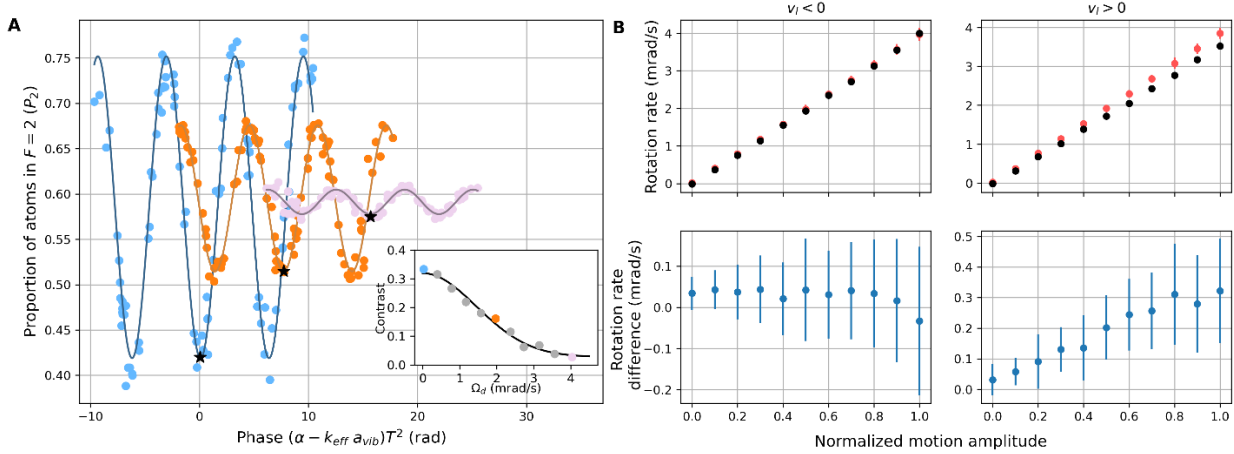

**Fig. 5. Measurement of dynamic rotation rates.** (**A**) Vibrations corrected phase scan of the atomic interference fringes for $\Omega_d = 0$ (blue), 2 (orange) and 4 (purple) mrad/s, where experimental dots are fitted with a sinusoidal function. Black star markers follow the shift of the fringe pattern due to the Coriolis acceleration. The contrast of the fringe patterns' sinusoidal fit is displayed in the inset as a function of the rotation rates amplitude $\Omega_d$, with colors referring to each fringe showed in the main plot, and fitted with a decaying exponential function (black line). (**B**) Dynamic rotation rates (top) measured by the classical (red) and atomic (black) gyroscope and difference between the two sensors (bottom), as a function of the loudspeakers amplitude of excitation for negative (left) and positive (right) launch velocities. Each dot is the result of an average over 200 shots (one fringe scan).

## Discussion

We have developed a dual atomic accelerometer-gyroscope and demonstrated long-term bias stability improvement of respectively 100-fold and 3-fold over our classical accelerometer and gyroscope in static conditions, corresponding to bias stabilities of $6 \times 10^{-7}$ m/s² and $4 \times 10^{-7}$ rad/s after 2 days of integration. Our magnetic field gradient-based method for launching the cold-atom cloud combined with the use of a single Raman beam offers a compact atom interferometer scheme. The implementation of a second horizontal Raman beam, which can be used to perform acceleration and rotation rate measurements along another axis in the future, allows the capability for self-calibration of the scale factor, with demonstrated scale factor stability of 700 ppm for rotation rate measurements. The hybridized sensor computes continuously both the vertical acceleration and horizontal rotation rate and embraces the advantages of both the short-term sensitivity of each classical sensors and the long-term stability of the quantum one. Dynamic rotation rates have also been measured up to 4 mrad/s level, demonstrating a linear scaling of the phase-shift with the rotation rate.

Multi-dimensional inertial measurement is required for many practical applications where highly stable and precise sensors must be used either for positioning in the absence of GNSS signal (*23*, *24*) or to compute the orientation of the carrier for metrology measurements (*25*, *26*). Currently used inertial measurement units, where the acceleration and rotation rates are measured along all axis of space, suffer from bias drift that limits the estimated position and attitude of the carrier. The method demonstrated in this work, based on the inherent stability of matter-wave inertial sensors, could be used to increase the self-reliance of classical inertial measurement units. The original technique for rotation rate measurements presented here, where the atoms are launched thanks to a pair of coils and interrogated in the diameter of a single Raman beam, goes with compactness, simplicity and scalability. This architecture could be extended to a full six-axis inertial measurement unit by adding a pair of coils and a Raman beam along the other horizontal axis. This way, the rotation rate along both horizontal axes could be measured by launching the atom cloud along the other respective horizontal direction and interrogating them with a vertical Raman, whereas vertical rotation rate could be obtained by launching the atoms along one horizontal direction and interrogating them along the other. The sensitivity of the vertical gyroscope could worsen due to the free-fall of the atoms through the waist of the horizontal Raman beam.

This original method of atom launching offers compactness but a relatively low scale factor stability of 700 ppm compared to the usual moving molasses launching technique that reaches the tens of ppm (*42*). A trade-off between the scale factor stability and the short-term sensitivity limitation could be found by increasing the launch velocity and reducing the interrogation time $T$. The acceleration long-term stability demonstrated here is sufficient for high accuracy inertial navigation system. However, in order to compete with the best inertial units, further developments are required to improve the scale factor and long-term bias stabilities respectively by at least three and two orders. In order to reduce the stabilization timescale of classical gyroscopes, work is currently under way to reach state-of-the-art atomic gyroscope's sensitivities (*29*) in a compact sensor by implementing large momentum transfer atom optics (*46*). The results presented here also show the ability to track the rotation rate in presence of an experimentally simulated on-board environment, with a limit set by the contrast loss induced by the finite temperature of the atomic cloud. Overcoming this limit can be done by reducing the interrogation time $T$, at the cost of a dramatic loss of sensitivity, decreasing the atom cloud temperature (*47*), at the expense of a loss of bandwidth, or by actively compensating the rotation of the retro-reflecting reference mirror (*44*). With these improvements, we could provide the first demonstration of a full operating cold-atom inertial measurement unit with performances overcoming the classical technologies.

## Materials and Methods
### Details on the experimental sequence
A 3D MOT of ~$10^8$ $^{87}$Rb atoms is formed in 300 ms at the intercept of 3 retroreflected gaussian laser beams with $1/e^2$ diameter equal to the vacuum chamber window's diameter (34 mm), allowing to maximize the capture efficiency of the trap with a total optical power of 30 mW per beam. The strength of the magnetic field gradient is 6.6 G/cm (1 G = $10^{-4}$ T) in the vertical direction. The last 40 ms of the MOT loading phase are used to shift the center position of the MOT along the East-West axis of the apparatus. This is done by applying a static horizontal bias magnetic field of 2 G, allowing to shift the magnetic field zero and thus the MOT's center position by 7.2 mm. Polarization gradient cooling during 8 ms follows the MOT in order to cool the atoms down to 2 µK. The atoms are then released from the trap and start to fall freely under gravity. Following the release, the atoms are horizontally launched using a 20 ms magnetic gradient pulse of 11.3 G/cm, that is combined with a horizontal bias magnetic field of ~110 mG that defines a quantification axis. 25 ms later, a 500 µs micro-wave pulse couples the internal state $|F = 2, m_F = +1\rangle$ to $|F = 1, m_F = 0\rangle$. The atoms that have not been transferred to the $F = 1$ state are cleared away with a 1 ms push beam. The horizontal bias magnetic field is then rotated in order to set a vertical quantification axis for the two-photon stimulated Raman transitions. The magnetic field is rotated adiabatically in 2 ms such that the spin of the atoms remains aligned with the magnetic field and the prepared state $|F = 1, m_F = 0\rangle$ is preserved (*48*). The two states of the atom interferometer $|F = 1, m_F = 0\rangle$ and $|F = 2, m_F = 0\rangle$ are coupled using a single vertical phase modulated Raman laser encoding the two frequencies required for the two-photon transition (*49*). Both frequencies are detuned by 956 MHz from the excited state $5^2P_{3/2}$ to avoid spontaneous emission. The Raman beam diameter $1/e^2$ has been precisely measured to be 20.2 mm. The interferometer is then realized with a combination of $\pi/2$ – $\pi$ – $\pi/2$ pulses equally separated by a duration $T$ and with the $\pi$-pulse, of duration equal to 10 µs, coinciding with the center of the Raman beam to minimize some systematic effects (*50*). After a total time of flight of 143 ms (~10 cm of free fall), the atoms are detected with a sequence of fluorescence pulses that allows to count the number of atoms in the hyperfine states $F = 2$ and $F = 1$.

**Shifting of the MOT center position and atom cloud's launching**
The same pair of horizontally aligned coils (named "launching coils" in Fig. 1.A.) is used for shifting the MOT's center position and launching the atoms. Each coil consists of 40 elliptical turns of 1 mm large copper wire wrapped directly around the vacuum chamber's arms. The elliptical shape (semi major-axis of 38.5 mm and semi-minor axis of 22.5 mm) is due to the form of the vacuum chamber and was chosen such that the atoms would experience the strongest magnetic field gradient possible. We have run numerical calculations demonstrating that this configuration creates a quasi-homogenous magnetic field and gradient respectively in Helmholtz and anti-Helmholtz configurations.

The shifting of the MOT's center occurs when the magnetic field zero's position created by the MOT's coils is modified due to the presence of an additional magnetic field. For instance, we use the pair of coils in a quasi-Helmholtz configuration where the same current runs inside both coils. Two cameras positioned at 45° from the East-West axis and 90° from one to each other's enable to image the MOT's position just before the optical molasses. We measured a linear shift of the MOT's position of 3.4 mm/G along the launch axis with negligible displacements along the two others.

This same pair of coils is also used for the atoms launching. Here, we create a magnetic field gradient by running the two coils in a quasi-anti-Helmholtz configuration. To swap between the MOT's position shifting and the atom cloud's launching, two electronic relays are implemented in order to invert the sign of the current in one coil. The sign of the launch velocity $\pm \vec{v}_l$ is alternated with another pair of electronic relays that reverses the sign of the current in both coils.

**Non-launched atoms selection**
After the cooling stage, the atoms are all in the $|F = 2\rangle$ hyperfine state. They are then launched horizontally with a magnetic field gradient pulse and selected with a one-photon micro-wave transition, with the aim to keep only the launched atoms in the state $|F = 2, m_F = +1\rangle$. This is done using the transition $|F = 2, m_F = +1\rangle \rightarrow |F = 1, m_F = 0\rangle$. However, because the hyperfine states $F = 1$ and $F = 2$ of the $^{87}$Rb have the same Zeeman splitting with Landé factors of opposite signs, this transition is degenerated with the other micro-wave transition $|F = 2, m_F = 0\rangle \rightarrow |F = 1, m_F = +1\rangle$. As the push beam only clears away the atoms in the state $|F = 2\rangle$, the non-launched atoms have therefore the possibility to participate to the atom interferometer and be detected.

The contribution of the non-launched atoms in the measured phase shift is yet very negligible for two reasons. First, they fall on the edge of the Raman beam (waist of 10.1 mm) due to the shift of the MOT's center position by 7.2 mm. Thus, they are only submitted to a small fraction of the required power necessary to perform the Mach-Zehnder atom interferometer. Second, the detection beam has a narrow waist of 3.3 mm and can be shifted horizontally (see Fig.1.C.), which allows to selectively detect the atoms in a specific spatial region. The atom cloud's spatial extension has been measured to be 2.5 mm after the optical molasses, resulting in a 5.5 mm wide cloud at the moment of the detection stage (150 ms of free-fall at a temperature of 2 µK). The center of the non-launched cloud being spaced by 12.2 mm from the center of the detection beam, we consider that a negligible fraction of the non-launched atoms is detected.

**Light-shifts removal for the launch velocity measurements**
The launch velocity analysis is performed by Raman spectroscopy using two-photon stimulated Raman transitions between the states $|F = 1, \vec{p}\rangle$ and $|F = 2, \vec{p} + \hbar\vec{k}_{eff}\rangle$. However, these states can respectively be coupled with the other neighbor atomic states $|F = 2, \vec{p} - \hbar\vec{k}_{eff}\rangle$ and $|F = 1, \vec{p} + 2\hbar\vec{k}_{eff}\rangle$, resulting in a two-photon light shift (TPLS) of the aimed transition's frequency (41). Here, we record the launch velocity by measuring the frequency difference $\Delta\nu$ between the two counterpropagating transitions $\pm\vec{k}_{eff}$. One-photon light shifts have no consequences on the measurement because they will move both atomic transitions by the same quantity. However, the two-photon light shifts are not equal for both atomic transitions, which results in a shift $\delta\omega_{TPLS}$ of $\Delta\nu$ that depends on the effective Rabi frequency $\Omega_{eff}$ of the coupling between $|F=1,\vec{p}\rangle$ and $|F=2,\vec{p}+\hbar\vec{k}_{eff}\rangle$, $\delta\omega_{TPLS} = -\Omega_{eff}^2\left[\frac{1}{4\omega_D}+\frac{1}{8\omega_D+16\omega_r}+\frac{1}{8\omega_D-16\omega_r}\right]$ where $\omega_D = \vec{k}_{eff}\cdot\vec{v}$ and $\omega_r = \frac{\hbar k_{eff}^2}{2m}$ are respectively the Doppler and recoil frequencies associated to the two-photon stimulated Raman transitions.

Figure 6 shows the measurements of the frequency difference $\Delta\nu$ for different values of $\Omega_{eff}$. This was experimentally done by changing the Raman laser power while adjusting the duration $\tau$ of the Raman pulse in order to get pulses which have the same total energy. It enables to maximize the visibility of each Raman spectrum because the condition $\Omega_{eff}\times\tau = \pi$ is always satisfied. However, the longer the Raman pulse lasts, the more selective in velocity the pulse gets which has for consequences to reduce the width of the transitions but also the signal-to-noise ratio. The effect of the TPLS is also reduced when increasing the Raman pulse duration. The velocity measurements presented in Fig. 2 were done by applying a 20 µs Raman pulse which is a good trade-off between pointing accuracy of the transition, signal-to-noise ratio and TPLS's amplitude.

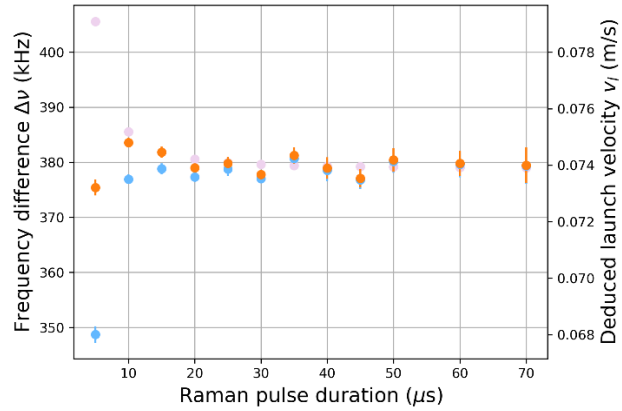

**Fig. 6. Two-photon light shift analysis.** Difference of frequency $\Delta\nu$ between the two transitions $\pm\vec{k}_{eff}$ without (blue) and with (orange) two-photons light shift correction (purple, plotted with an offset for clarity) as a function of the Raman pulse duration. The horizontal Raman laser intensity and pulse duration are changed in order to satisfy the condition $\Omega_{eff}\times\tau = \pi$. The launch velocity scale is deduced ($v_l = \Delta\nu\times\pi/k_{eff}$) showing a possible 10% mismatch on the real value if not corrected from the two-photons light shift.

**Ground vibration removal**
Ground vibrations can propagate to the retro-reflecting mirror. As the atom interferometer measures the acceleration of the atoms compared to the retroreflecting mirror, this produces phase noise in the atomic phase shift. To remove it as accurately as possible, we record the acceleration sensed by the retro-reflecting mirror with the force-balanced accelerometer $a_{class}$ during the atom interferometry stage and convolute it with the atom interferometer's sensitivity function (51):

$$a_{conv,at} = \int_{-T}^{T} a_{class}(t_\pi + t) \times h_{at}(t)\, dt \qquad (6)$$

where $t_\pi$ is the instant of the second laser pulse of the atom interferometer and $h_{at}$ is the acceleration sensitivity function. For a Mach-Zehnder geometry, this is a triangle-like function that is maximum at the instant of the second pulse:

$$h_{at}(t) = \begin{cases} \dfrac{T+t}{T^2} & \text{if } t \in [-T, 0] \\ \dfrac{T-t}{T^2} & \text{if } t \in [0, T] \end{cases} \qquad (7)$$

The ground vibrations noise $\Delta a_{vib}$ is then removed by adding the term $\Delta a_{corr}/8v_l$ to the rotation rate measurement in equation (3) with $\Delta a_{corr} = a_{conv,at}^{+k,+v} + a_{conv,at}^{-k,+v} - \left(a_{conv,at}^{+k,-v} + a_{conv,at}^{-k,-v}\right)$ corresponding to the four configurations of the atom interferometer described in Figure 3 ($\pm \vec{k}_{eff}, \pm \vec{v}_l$). Figure 7 exhibits the atomic interferometer output $\alpha_{\pm k, \pm v}$, in acceleration units, and the convoluted classical accelerometer output $a_{conv,at}^{\pm k, \pm v}$. We observe a clear correlation between the two signals, meaning that the atomic gyroscope sensitivity is limited by the ground vibrations noise. The correlations are however not perfect, which underlines the limit of the vibration noise removal using this method.

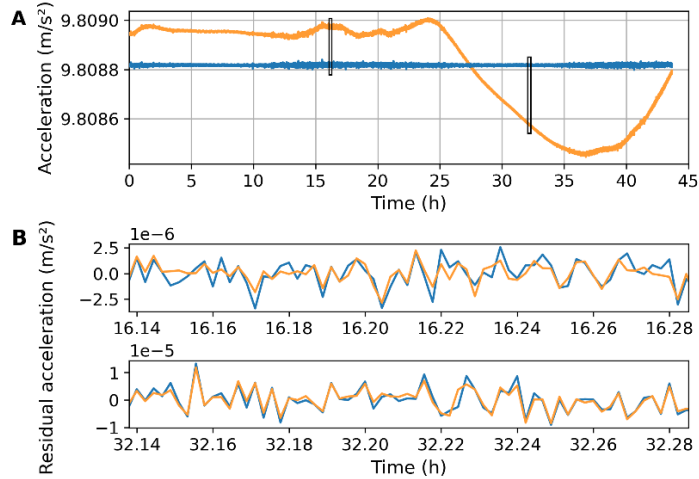

**Figure 7. Ground vibrations sensing by the classical and atomic accelerometers. (A)** The full temporal track of the atomic interferometer output $\alpha_{\pm k, \pm v}$ (blue), in acceleration units, and the convoluted classical accelerometer output $a_{conv,at}^{\pm k, \pm v}$ (orange) are plotted. **(B)** Zooms over small timescales are performed, where both acceleration signals subtracted from their mean value are compared to highlight their correlation.

**Estimation of the wave front aberrations phase shift**
In our atom interferometer configuration, the atoms travel in a plane perpendicular to the Raman beam propagation direction between the three laser pulses. For each Raman light pulse, the effective Raman phase of the upward and downward laser beams $\varphi_i = \varphi_i^{up} - \varphi_i^{down}$ ($i = 1, 2, 3$) is carved onto the phase of the atoms wave-function. As the Raman light pulses are not plane waves, the carved phase therefore depends on the atoms position within the beam profile. For an atom cloud launched with a velocity $\pm \vec{v}_l$, the phase shift of the atom interferometer can be computed by summing the laser phase at each pulse $\varphi_1^\pm - 2\varphi_2^\pm + \varphi_3^\pm$ (see Figure 8), where the subscript $\pm$ is associated to the $+\vec{v}_l$ and $-\vec{v}_l$ atom interferometers.

For the following calculation, we neglect the diffraction of the wave-front and the expansion of the atomic cloud during the interferometer. If the two interferometers are perfectly symmetric, the phase shift due to the wave-front aberrations acquired in one interferometer will cancel out with the phase shift acquired in the other because the rotation rate is calculated as a difference of these two configurations (see equation (3)). Thus, even a very distorted wave front has no impact on the rotation rate measurement under the assumptions made. However, if the two interferometer configurations are not symmetric, an additional phase shift associated to the wave-front aberration appears and scales with the position asymmetry $\delta x$. In the schematic example of Figure 8, this additional phase shift corresponds to $\varphi_1^+ - \varphi_3^-$.

We consider an imperfect Raman wave front that has a deformation modeled by a polynomial of order $k$, making the laser phase carved on the atoms at each pulse $i = 1, 2, 3$ to be $\varphi_{k,i}^\pm = A_k x_i^k$. The additional phase shift due to wave front aberrations is $\Delta\Phi_{ab,k}^\pm = \varphi_{k,1}^\pm - 2\varphi_{k,2}^\pm + \varphi_{k,3}^\pm$ for each interferometer, leading to an error on the rotation rate measurement $\Delta\Omega_{ab,k} = (\Delta\Phi_{ab,k}^+ - \Delta\Phi_{ab,k}^-)/4v_l k_{eff} T^2$.

For a deformation of order $k = 2$, this error is null because of the symmetry of the aberration compared to the Raman beam center. More generally, this is true for every even order of $k$. Yet, this is not true if we consider a deformation of odd order of $k$. For instance, if $k = 3$, the aberration phase shift of one interferometer is $\Delta\Phi_{ab,3}^\pm = \pm 6A_3 v_l^2 T^2 \delta x$ and leads to an error on the rotation rate measurement:

$$\Delta\Omega_{ab,3} = \frac{3A_3 v_l}{k_{eff}} \delta x \qquad (8)$$

where $A_3 = \frac{2\pi/\lambda \times 2\, OQ}{w^3}$ depends on the optical quality $OQ$ of the window, $w = 10.1$ mm the waist of the Raman beam and $\delta x \approx 0.6$ mm is the estimated dissymmetry. In our experimental setup, the main source of Raman wave front error is the viewport of the vacuum chamber through which the Raman beam goes twice. A transmitted wave front error of the viewport of $\lambda/6$ peak to valley over $2w$, which is compatible with the expected quality of the optic, can lead to the difference of $1.5 \times 10^{-5}$ rad/s between the measurement presented in this work and the expected value of the Earth rotation projected on the measurement axe.

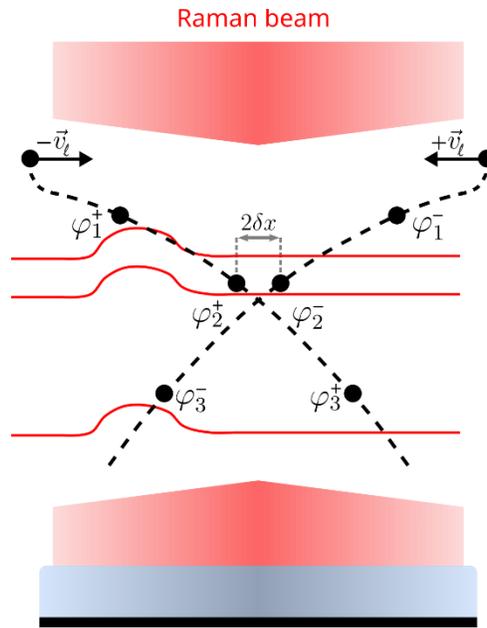

**Figure 8. Wave front aberrations phase-shift model.** The mid-point trajectory of each ($\pm\vec{v}_l$) atom interferometer is drawn with the positions of the corresponding pulses $i = 1, 2, 3$. The wave front of the Raman beam is represented with a deformation that propagates along its propagation direction. A phase $\varphi_i^{\pm}$ that depends on the atoms position due to the deformation is imprinted on the atomic wave-packet at each pulse and does not cancel out if the symmetry of the atom interferometer is broken ($\delta x \neq 0$).

**Funding:** We acknowledge the financial support of Agence Innovation Défense (AID), Ecole Doctorale Ondes et Matières (EDOM) and ONERA through the research project CHAMOIS.

**Author contributions:**
  Conceptualization of the experimental apparatus: C.S., Y.B., M.C., N.Z., C.B. and A.Br.
  Build the experimental setup: C.S., Y.B., M.C., C.B., A. Br.
  Conduct the experiments: C.S.
  Data analysis: C.S., Y.B. and M.C.
  Investigation: C.S., Y.B., M.C., S.D., A. Bo., S.S., A.Br.
  Writing – original draft: C.S., M.C. and S.D.
  Writing – review & editing: All the authors

**Competing interests:** Authors declare that they have no competing interests.

**Data and materials availability:** All data needed to evaluate the conclusions in the paper are present in the paper and/or the Supplementary Materials. Raw data are available at the following repository: https://zenodo.org/records/11241234